\documentclass[12pt]{article}
\usepackage{amsmath}
\usepackage{amsfonts}
\usepackage{graphicx}
\usepackage[spanish]{babel}

\begin{document}

\title{Aut\'omatas celulares elementales aplicados a la encriptaci\'on de datos}
\author{\textbf{Elena Villarreal Zapata y Francisco Cruz Ordaz Salazar}\\Universidad Polit\'ecnica de San Luis Potos\'i}

\maketitle

\begin{abstract}

Para el cifrado de datos suele ser necesaria una llave como base, por lo que es indispensable tener una que sea robusta y confiable, para as\'i evitar el acceso de terceros a la informaci\'on cifrada. 
Esto requiere un generador de n\'umeros pseudo-aleatorios que proporcionar\'a dicha llave, por lo que se propone trabajar con aut\'omatas celulares auxili\'andose con \emph{Mathematica}\index{Mathematica}, para revisar qu\'e reglas, y a qu\'e nivel, son pseudo-aleatorias. Este proyecto se centra en la revisi\'on de posibles reglas pseudo-aleatorias, analizando sus caracter\'isticas detalladamente y someti\'endolas a un conjunto de pruebas de aleatoriedad con el fin de conocer cuales de ellas nos permitir\'an obtener los n\'umeros pseudo-aleatorios\index{pseudoaleatoriedad} que conformar\'an la llave para el cifrado de datos.


\end{abstract}

\section{Introducci\'on}

Esta investigaci\'on es parte complementaria de un proyecto que se est\'a trabajando en la Universidad Polit\'ecnica de San Luis Potos\'i, en el que se pretende desarrollar un sistema de encriptaci\'on\index{encriptaci\'on} de datos basado en aut\'omatas celulares\index{aut\'omatas celulares}.
Como parte del proyecto inicial se debe comprobar que se est\'a trabajando con reglas pseudo-aleatorias, ya que de comenzar a trabajar con reglas al azar se corre el riesgo de generar un cifrado que puede ser \emph{hackeado} f\'acilmente. 
Por tanto, se buscan reglas de aut\'omatas celulares que tengan un comportamiento pseudo-aleatorio, para despu\'es generar secuencias con cada una y probar su aleatoriedad\index{aleatoriedad}.

\section{Antecedentes}
Los aut\'omatas celulares fueron inventados a fines de los a\~nos cuarenta por Stanislaw Ulam y John von Neumann, quienes realizaron trabajos para crear un sistema que se replicara a s\'i mismo a partir de una abstracci\'on matem\'atica.
A\~nos despu\'es, Wiener\index{Norbert Wiener} y Rosenblueth\index{Arturo Rosenblueth} desarrollaron un modelo de aut\'omatas celulares que pretend\'ia describir matem\'aticamente la conducci\'on de impulsos en sistemas cardiacos. En los sesentas se empezaron a estudiar como un tipo de sistemas din\'amicos, y para los setenta aparece el Juego de la Vida\index{Juego de la Vida}.
Este fue inventado por John Conway\index{John Conway} y consist\'ia en una colecci\'on de celdas las cuales, basadas en reglas matem\'aticas, pod\'ian vivir, morir o mutiplicarse, todo esto dependiendo de las condiciones iniciales \cite{martin}.
En 1983, Stephen Wolfram public\'o algunos escritos sobre una clase de aut\'omatas que el llamaba aut\'omatas celulares elementales y sobre su comportamiento y las reglas que los defin\'ian.
Para el 2002, Wolfram\index{Stephen Wolfram} public\'o su libro A New Kind of Science\index{A New Kind of Science} \cite{wolfram} en el cual explica ampliamente sobre ellos, su trabajo  y su importancia en todas las ramas de la ciencia.
En cuanto a la encriptaci\'on, Olu Lafe \index{Olu Lafe} \cite{lafe} nos explica que existen un numero de patentes dadas y literatura sobre ello que incluye los trabajos de Wolfram (1985) \cite{swolfram}, Delahaye\cite{delahaye}\index{Jean-Paul Delahaye} (1991), Guan \index{Puhua Guan}\cite{guan} (1987) y Gutowitz \cite{gutowitz}\index{H.A. Gutowitz} (1994).
En los cuales, Wolfram hace uso de la regla 30 de los aut\'omatas celulares para generar n\'umeros pseudo-aleatorios; Guan usa un sistema din\'amico invertible; Gutowitz (U.S. Patent 5,365,589) usa sistemas din\'amicos irreversibles; y Lafe (U.S. Patent 5,677,956 el 14 de octubre de 1997) utiliza operaciones simples de transformaci\'on, lo cual implica una enorme biblioteca de llaves o c\'odigos criptogr\'aficos derivados de los aut\'omatas celulares\cite{delahaye,gutowitz,gutowitz2}.

\section{Aut\'omatas Celulares}
Un aut\'omata celular, en su versi\'on m\'as simple, es una l\'inea unidimensional de sitios o celdas, donde cada una es blanca o negra. El color o estado de esta celda puede cambiar conforme al tiempo.
Con cada paso discreto de tiempo, las celdas se actualizan (ya sea para mantener o cambiar su color previo) de acuerdo a la funci\'on de su estado anterior y al de las dos celdas vecinas a ella (una por el lado izquierdo y otra por el lado derecho).
Existen adem\'as, otros espacios disponibles de aut\'omatas celulares, donde se consideran m\'as par\'ametros como lo son el n\'umero de estados en las celdas, vecindarios mayores, plantillas m\'as amplias y dimensiones adicionales, colores, entre otros.

\subsection{Reglas}
A las condiciones de vecindad de un aut\'omata celular se le conoce como ``regla". 
Existen 256 ($2^8$) reglas para los aut\'omatas celulares con un estado binario variable (0,1) y una vecindad de 1 con longitud de tres. Cada una de ellas est\'a especificada por un c\'odigo decimal obtenido a partir de las ocho permutaciones para la vecindad 1 en orden descendiente y los leemos como un código binario de ocho d\'igitos, lo cual nos da el n\'umero de la regla. La regla 30 \index{regla 30}, por ejemplo, est\'a definida  por la configuraci\'on dada en la figura 1. N\'otese que la secuencia 00011110 es la representaci\'on binaria del n\'umero 30.

\begin{figure}[h!]
\begin{center}\label{elena}
\includegraphics{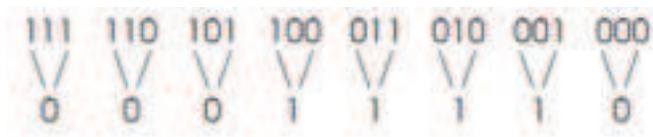}
\caption{Representaci\'on binaria de la regla 30.}
\end{center}
\end{figure}
 
Stephen Wolfram propone un esquema de clasificaci\'on\index{clasificaci\'on de Wolfram}, el cual divide las reglas de aut\'omatas celulares en cuatro categor\'ias de acuerdo a sus evoluciones a partir de una condici\'on inicial ``desordenada" o aleatoria. La clase 1, tambi\'en conocida como de tipo fijo, la cual evoluciona r\'apidamente a un estado estable y homog\'eneo en el que todos los sitios tienen el mismo valor y cualquier aleatoriedad en el patr\'on inicial desaparece; la clase 2, tambi\'en conocida como de tipo peri\'odico, en la cual se repite un mismo patr\'on como un bucle donde su evoluci\'on es a gran velocidad y cualquier aleatoriedad en el patr\'on inicial solo dejar\'ia restos que complementar\'ian el bucle; la clase 3, tambi\'en conocida como de tipo ca\'otico o pseudo-aleatorio, en donde su evoluci\'on conduce a un patr\'on ca\'otico donde cualquier estructura estable es r\'apidamente destruida por el ruido circundante y los cambios tienden a extenderse de manera indefinida; y la clase 4, de tipo complejo, la cual presenta comportamientos tanto de la clase 2 y 3 y suelen presentar una evoluci\'on m\'as lenta.
Teniendo una condici\'on inicial simple, existen 13 reglas de aut\'omatas celulares en las clases 3 y 4 calculadas en \cite{zenil}\index{Hector Zenil}, que son las siguientes: 30, 45, 75, 79, 86, 89, 101, 110, 124, 135, 137, 149, 193. 
\'Estas se ampl\'ian a 38 reglas de clase 3 si se tienen condiciones iniciales aleatorias \cite{swolfram}, las cuales son las siguientes: 18, 22, 30, 45, 54, 60, 73, 75, 86, 89, 90, 101, 102, 105, 106, 109, 110, 120, 122, 124, 126, 129, 135, 137, 146, 147, 149, 150, 151, 153, 161, 165, 169, 182, 183, 193, 195 y 225. 
Seg\'un un estudio realizado en Brasil \cite{mattos}, las reglas de clase 3 pueden ser clasificadas en cuatro distintas subclases: Dep\'osito Aleatorio\index{Dep\'osito Aleatorio} (Declaraci\'on), representada por las siglas RD; Percolaci\'on Dirigida\index{Percolaci\'on Dirigida}, representada por las siglas DP; Percolaci\'on Compacta Dirigida\index{Percolaci\'on Compacta Dirigida}, de siglas CDP; y aut\'omatas celulares Domany-Kinzel\index{aut\'omatas celulares Domany-Kinzel}, de siglas DKCA y donde pueden ser sim\'etricos o asim\'etricos. 
Siendo que las reglas de clase 3 presentan comportamientos ca\'oticos y pseudo-aleatorios, se eligieron cuatro reglas. La regla 30 perteneciente a la subcategor\'ia RD; la regla 54 perteneciente a la subcategor\'ia DKCA (asim\'etrica); la regla 73 perteneciente a la subcategor\'ia CDP; y la regla 110 perteneciente a la subcategor\'ia  DP y DKCA (sim\'etrica).

\section{Pseudo-aleatoriedad}
La necesidad de obtener n\'umeros aleatorios y pseudo-aleatorios se plantea en muchas aplicaciones criptogr\'aficas, pues se emplean llaves que deben ser generadas con dichas caracter\'isticas. Por ejemplo, para cantidades auxiliares usadas en generaci\'on de firmas digitales, \'o para generar desaf\'ios en autentificaci\'on de protocolos. 
El Instituto Nacional de Est\'andares y Tecnolog\'ia (NIST)\index{suite de tests NIST} proporciona un conjunto de pruebas estad\'isticas de aleatoriedad y considera que estos procedimientos son \'utiles en la detecci\'on de desviaciones de una secuencia binaria en la aleatoriedad \cite{rukhin}\index{Andrew Rukhin}.
Existen dos tipos b\'asicos de generadores usados para producir secuencias aleatorias: Generadores de N\'umeros Aleatorios (RNGs) y Generadores de N\'umeros Pseudo-Aleatorios (PRNGs)\index{generador de n\'umeros pseudo-aleatorios}. Para aplicaciones criptogr\'aficas\index{criptograf\'ia}, ambos tipos de generadores producen un flujo de ceros y unos que pueden ser divididos en sub-flujos \'o bloques de n\'umeros aleatorios. Nuestro inter\'es est\'a en la revisi\'on de un generador tipo PRNGs, en este caso, si la semilla (l\'inea inicial) es desconocida, en el paso siguiente el n\'umero producido en la secuencia debe ser impredecible a pesar de todo conocimiento de n\'umeros aleatorios anteriores en la secuencia. Esta propiedad se conoce como imprevisibilidad siguiente, y es lo que se presume que obtenemos mediante aut\'omatas celulares de clase 3.
El conjunto de pruebas de NIST es un paquete estad\'istico que consiste en 15 pruebas que se desarrollaron para probar la aleatoriedad de (arbitrariamente largas) secuencias binarias producidas por hardware y software basado en generadores criptogr\'aficos de n\'umeros aleatorios o pseudo-aleatorios. Dichas pruebas se enfocan en diversos tipos de no aleatoriedad que pueden existir en una secuencia. Las 15 puebas son:

\begin{itemize}
\item Prueba de frecuencia (Monobit)\index{monobit}.
\subitem Esta prueba mide la proporci\'on de ceros y unos de toda una secuencia.
\item Prueba de frecuencia dentro de un bloque.
\subitem Esta prueba mide la proporci\'on de unos dentro de un bloque de $M$ bits.
\item Prueba de corridas.
\subitem Esta prueba mide el total de corridas en una secuencia, donde una corrida es una secuencia interrumpida de bits id\'enticos.
\item Prueba de la m\'as larga corrida de unos en un bloque.
\subitem Esta prueba mide la corrida m\'as larga de unos dentro de un bloque de $M$ bits.
\item Prueba de rango de la matriz binaria.
\subitem Esta prueba mide el rango de sub-matrices disjuntas de toda la secuencia.
\item Prueba de la transformada discreta de Fourier (Espectral)\index{transformada de Fourier}.
\subitem Esta prueba mide las alturas de los picos en las transformadas discretas de Fourier de las secuencias.
\item Prueba de la no acumulaci\'on de coincidencia de plantilla.
\subitem Esta prueba mide el n\'umero de ocurrencias de cadenas destino pre-especificadas. Una ventana de m bits es usada para buscar un patr\'on espec\'ifico de m bits.
\item Prueba de acumulaci\'on de coincidencia de plantilla.
\subitem Esta prueba tambi\'en mide el n\'umero de ocurrencias de cadenas destino pre-especificadas. La diferencia con la prueba anterior reside en la acci\'on realizada al encontrar un patr\'on.
\item Prueba de Estad\'istica Universal de Maurer.
\subitem Esta prueba mide el n\'umero de bits entre los patrones de juego (una medida que est\'a relacionada con la longitud de una secuencia comprimida).
\item Prueba de complejidad lineal.
\subitem Esta prueba mide la longitud de un Registro de Desplazamiento con Retroalimentaci\'on Lineal (LFSR). Una baja longitud LFSR implica no aleatoriedad.
\item Prueba de serie.
\subitem Esta prueba mide la frecuencia de todos los posibles patrones de m bits acumulados a trav\'es de la secuencia completa.
\item Prueba de entrop\'ia aproximada.
\subitem Esta prueba tiene el mismo enfoque que la anterior, con el prop\'osito de comparar la frecuencia de bloques acumulados de dos consecutivas/adyacentes longitudes ($m$ y $m+1$).
\item Prueba de sumas acumulativas.
\subitem Esta prueba mide la excursi\'on m\'axima (desde cero) del paseo aleatorio definido por la suma acumulada de ajustados (-1, +1) d\'igitos en la secuencia.
\item Prueba de excursiones aleatorias.
\subitem Esta prueba mide el n\'umero de ciclos teniendo exactamente k visitas en una suma acumulativa de un paseo aleatorio.
\item Prueba variante de excursiones aleatorias.
\subitem Esta prueba mide el total de veces que un estado particular es visitado (es decir, se produce) en una suma acumulada de un paseo aleatorio.
\end{itemize}

\section{Metodolog\'ia}

Primeramente se recopil\'o informaci\'on sobre las clases que propone Wolfram para clasificar las reglas del aut\'omata celular. Fue con esta recopilaci\'on, que se encontr\'o que exist\'ian subcategor\'ias propuestas dentro de la clase 3. Y, al encontrar estas subcategor\'ias, se decidi\'o realizar pruebas de aleatoriedad a una regla por divisi\'on, como se mencion\'o anteriormente. 

\begin{table}[!h]
\label{tabla1}
\caption{Resultados de las pruebas aplicadas a reglas representantes de cada una de las 4 clases de Wolfram (A significa que la prueba fue aprobada y R que fue reprobada).}
\begin{center}
\begin{tabular}{|l|l|l|l|l|}
  \hline
Prueba& R30& R54& R73& R110\\
  \hline
Frecuencia (Monobit)&A&R&R&R\\
Frecuencia dentro de un bloque&A&R&R&R\\
Corridas&R&R&R&R\\
M\'as larga corrida de unos en un bloque&A&R&R&R\\
Rango de la matriz binaria&A&R&A&R\\
Transformada discreta de Fourier (Espectral)&R&R&R&R\\
No acumulaci\'on de coincidencia de plantilla&A&A&A&A\\
Acumulaci\'on de coincidencia de plantilla&A&A&R&R\\
Estad\'istica Universal de Maurer&A&A&R&R\\
Complejidad lineal&A&R&A&R\\
Serie&A&R&R&R\\
Entrop\'ia aproximada&A&R&R&R\\
Sumas acumulativas&A&R&R&R\\
Excursiones aleatorias&A&R&R&R\\
Variante de excursiones aleatorias&A&R&R&R\\
\hline
\end{tabular}
\end{center}
\label{default}
\end{table}

Por tanto, para cada una de las reglas elegidas, se generaron mediante \emph{Mathematica} 1000 archivos con 10000 datos. Estos 10000 datos son conformados a partir de una `cadena inicial de 100 caracteres, la cual es generada aleatoriamente y se compone \'unicamente de 0s y 1s. 
Despu\'es de generar los archivos para cada regla, estos se juntaron en un solo archivo, que posteriormente se analizar\'ia mediante la Suite de Pruebas de la NIST.
Al finalizar el an\'alisis de cada archivo final (uno por regla), se obtuvo un archivo con los resultados del an\'alisis, lo cual nos permite ver si la regla tiene o no caracter\'isticas que la avalen como pseudo-aleatorias o no.  En el Cuadro 1, podemos ver una comparaci\'on de las reglas y su pase en cada una de las pruebas. 

Los resultados y calificaciones de la tabla se obtuvieron despu\'es de realizar varias veces el procedimiento de generaci\'on y prueba de datos y promediar los resultados por prueba estad\'istica y por intento.

\section{Conclusiones}
Como podemos ver, la regla que m\'as propiedades de pseudo-aleatoriedad presenta es la regla 30, por lo que podemos concluir que se puede considerar que es pseudo-aleatoria. Es importante notar que las dos pruebas que reprueba no se les considera que afecten a los resultados, puesto que se not\'o que sus reprobaciones son debido a que las corridas son consideradas perfectas, lo cual es poco probable en un generador de n\'umeros pseudo-aleatorios. 
Por el contrario, la enorme falta de propiedades b\'asicas de aleatoriedad en las otras reglas, nos permite pensar que es posible que solo las reglas de clase 3 que pertenezcan a la subcategor\'ia RD sean las que presenten pseudo-aleatoriedad.
Se contin\'ua realizando pruebas nuevamente, con reglas distintas a las elegidas, para comprobar si es la subcategor\'ia o si solamente fue una coincidencia entre las reglas elegidas de cada subcategor\'ia que solo aquella perteneciente a la subcategor\'ia RD sea pseudo-aleatoria.
Consideramos que despu\'es de realizar estas pruebas se podr\'ia continuar con el trabajo enfoc\'andose a la encriptaci\'on y recomendamos que se pruebe cada una de las reglas pseudo-aleatorias encontradas, como llave de un sistema simple de encriptaci\'on y, posteriormente, en uno m\'as complejo para verificar el funcionamiento de las mismas como llaves y su utilidad.


\begin{thebibliography}{00}

\bibitem{delahaye} J-P. Delahaye, Les Automates, \emph{Pour La Science,} pp.126--134, 1991.
\bibitem{gutowitz} H. A. Gutowitz, Cellular Automata: Theory and Experiment; proceedings of an interdisciplinary workshop, Editor. vol. 45, \emph{Physica D,} 1990.
\bibitem{gutowitz2} H. A. Gutowitz,  Artificial Life Simulators and Their Applications, \emph{DRET Technical Report,} 1994.
\bibitem{rukhin} A.L. Rukhin, J. Soto, J. Nechvatal, M. Smid, E. Barker, S. Leigh, M. Levenson, M. Vangel, D. Banks, A. Heckert, J. Dray, S. Vo, A Statistical Test Suite for Random and Pseudorandom Number Generators for Cryptographic Applications, \emph{NIST Special Publication 800-22 A}, (2001) revised 2010.
\bibitem{lafe} O. Lafe, Cellular Automata Transforms, \emph{Kluwer Academia Publishers,} 2000.
\bibitem{martin} E. Martin, John Conway's Game of Life. Recuperado el Julio de 2010, de http://www.bitstorm.org/gameoflife/
\bibitem{mattos} T.G. Mattos y J. G. Moreira. Universality Classes of Chaotic Cellular Automata, \emph{Brazilian Journal of Physics,} vol. 34, nœm. 02A, pp. 448--451, 2004.
\bibitem{guan} P. Guan, Cellular automaton public-key cryptosystem, \emph{Complex Systems,} vol. 1, pp. 51--57, 1987.
\bibitem{swolfram} S. Wolfram, Theory and Applications of Cellular Automata, \emph{Rev. Mod. Phys.} 55, 601, 1983.
\bibitem{wolfram} S. Wolfram, \emph{A New Kind of Science,} Wolfram Media Inc., 2002.
\bibitem{zenil} H. Zenil, Compression-based Investigation of the Dynamical Properties of Cellular Automata and Other Systems, \emph{Complex Systems,} 19(1), pages 1-28, 2010. 

\end{thebibliography}
\end{document}